\documentclass[doublecol]{epl2}
\usepackage{amsmath,amssymb,graphicx}
\newcommand{\beq}{\begin{equation}}
\newcommand{\eeq}{\end{equation}}
\newcommand{\bea}{\begin{eqnarray}}
\newcommand{\eea}{\end{eqnarray}}
\newcommand{\ovl}[1]{\overline{#1}}

\title{Identifying short motifs by means of extreme value analysis}
\shorttitle{Identifying short motifs by means of extreme value analysis} 

\author{D. Bianchi\inst{1} \and B. Tirozzi\inst{1}}
\shortauthor{D. Bianchi and B. Tirozzi}

\institute{
  \inst{1} Dipartimento di Fisica, Universit\`a di Roma ``La Sapienza'', p.le Aldo Moro 2, 00185 Rome (Italy)
} 

\abstract{The problem of detecting a binding site -- a substring of DNA where
transcription factors attach -- on a long DNA sequence requires the recognition
of a small pattern in a large background. For short binding sites, the matching
probability can display large fluctuations from one putative binding site to another.
Here we use a self-consistent statistical procedure that accounts correctly for the
large deviations of the matching probability to predict the location of short binding
sites. We apply it in two distinct situations: (a) the detection of the binding sites
for three specific transcription factors on a set of 134 estrogen-regulated genes;
(b) the identification, in a set of 138 possible transcription factors, of the ones binding
a specific set of nine genes. In both instances, experimental findings are reproduced
(when available) and the number of false positives is significantly reduced with
respect to the other methods commonly employed.}

\begin{document}

\maketitle

\section{Introduction}

Understanding the regulation of gene expression, {\it i.e.} the
cellular process that controls the amount and timing of
appearance of the functional product of a gene, is a challenging
task. The expression of a gene is controlled by proteins called
transcription factors, which bind to short segments of DNA called
binding sites (BSs). BSs are located on long strings of DNA of
about 2000 nucleotides (the {\it promoters}), upstream of genes.
The problem of identifying BSs clearly plays a central  role for
elucidating the mechanics of gene regulation. The detection of BSs
can be carried out experimentally by several high-throughput
techniques (see e.g. \cite{exp}), though still at very high cost.
From such measurements it is possible to infer the frequency with
which every nucleotide (\texttt{A}, \texttt{C}, \texttt{G} or
\texttt{T}) appears in the BS of a given transcription factor.
These data ultimately represent the binding specificity of
transcription factors and are usually encoded in the so-called
Position-Specific Frequency Matrices (PSFMs) which are catalogued
in {\it e.g.} the JASPAR \cite{jaspar} and TransFac
\cite{transfac} databases. The entry $f_{ij}$ of a PSFM gives the
frequency with which nucleotide $j\in\{\texttt{A}, \texttt{C},
\texttt{G},\texttt{T}\}$ appears in position
$i\in\{1,\ldots,\ell\}$ on the BS ($\ell$ denoting its length) of
a given transcription factor, with $\sum_j f_{ij}=1$. Thousands of
such matrices are available today, covering BSs for many different
transcritption factors.
Developing effective computational methods to predict the position
of BSs on the promoter from the known PSFMs would produce a
crucial advantage in terms of identifying new BSs and improving
the characterization of binding specificity. From a theorist's
perspective, the question in somewhat simplified terms is the
following: given a long string of DNA (the promoter), which short
substring is the best putative BS according to the experimental
PSFM?

In order to tackle this issue, several methods have been developed
and are presently used \cite{pavesi,lass,sandve}. Most of them assume a
Markovian model as the underlying string generator and consist in
(i) a maximum-likelihood procedure to identify candidate substrings
on the promoter, and (ii) a statistical test to evaluate the
significance of the results against a benchmark in which
log-likelihoods are Gaussian-distributed. For short BSs this second
step is particularly delicate because their log-likelihoods are
expressed as sums of contributions coming from single nucleotides
treated independently. Therefore they can not be approximated
by Gaussian random variables \cite{bailey}, since the number of
terms in the sum is too small. Indeed we show below that the
probability distribution function (pdf) of the maximum of the
log-likelihoods for short BSs is rarely the extreme value
distribution for a Gaussian random variable. The prediction of short
BSs thus requires a more precise method that is able to account
correctly for large deviations in evaluating their statistical
significance.

In this work we apply the standard approach used in statistics
to evaluate the distribution of the maximum, {\it i.e.} extreme-value
theory and particularly the Peak-over-Threshold (POT) method, to
estimate the statistical significance of putative BSs. This
technique has been employed in financial analysis
\cite{embr} and meteorology (see e.g. \cite{meteo}). We firstly test our method
by identifying, among 138 transcription factors listed in JASPAR,
the ones binding a specific set of skeletal muscle specific genes, reproducing experimental
results with a marked reduction of false positives in comparison with
other computational methods \cite{def,tomovic}.
Subsequently, we apply it to the detection of a BS that is
widely studied experimentally, that is ERE \cite{ere2} (estrogen
responsive element, $\ell=13$), and of two other BSs that are
believed to be functionally related to ERE (called AP2 and C/EBP,
both $\ell=12$) on a data set of 134 promoters for genes whose
expression is altered upon treatment with an estrogen-sensitive
growth factor \cite{ere1}.

\section{Setup}

The basic setting of probabilistic schemes is in general terms as
follows. Consider a string of length $L$ drawn from a finite
alphabet $\mathcal{A}$. We assume that it can be divided in two
parts: a {\it background} consisting of $L-\ell$ letters and a {\it
motif} of length $\ell$ (the BS). These are produced in general by
different stochastic models $P_b$ and $P_m$ (the latter encoded by
the PSFM, in the case discussed above). Neglecting all correlations,
the probability of observing a certain sequence $S_k=\{a_1,\ldots,a_L\}$ of length $L$
including a motif that starts at location $k+1$ is simply \bea
P(S_k)&=&\underbrace{\prod_{i=1}^{k}
P_b(a_i,i)}_{\text{background}}\underbrace{\prod_{i=k+1}^{k+\ell}
    P_m(a_i,i)}_{\text{motif}}\underbrace{\prod_{i=k+\ell+1}^{L}P_b(a_i,i)}_{\text{background}}\nonumber\\
    &=&\prod_{i=1}^L P_b(a_i,i)\prod_{i=k+1}^{k+\ell}\frac{P_m(a_i,i)}{P_b(a_i,i)}\label{uno}
\eea where $P_b(a_i,i)$ (resp. $P_m(a_i,i)$) represents the
probability to observe letter $a_i\in\mathcal{A}$ in position $i$ in
the background (resp. the motif). It is clear that the motif can
be identified as the substring that maximizes the second factor in
the right-hand side of (\ref{uno}), or equivalently its logarithm,
{\it i.e.} \beq\label{two} W_k=\sum_{i=k+1}^{k+\ell}\left[\log
P_m(a_i,i)-\log P_b(a_i,i)\right] \eeq since larger values of $W_k$
suggest that the string starting at $k+1$ is more likely to be a
motif than a common substring. $W_k$ is called the {\it score} of
the substring. Moving $k$ along the string, one can then compute
$L-\ell+1$ scores, one for each substring of length $\ell$, and
select the one with the highest score as the most likely motif.
We shall henceforth denote by $k^\star+1$ the starting locus of
the score-maximizing substring. 
Note that multiple maxima may occur.

\section{Statistical significance}

\begin{figure}
\begin{center}
\includegraphics[width=7cm]{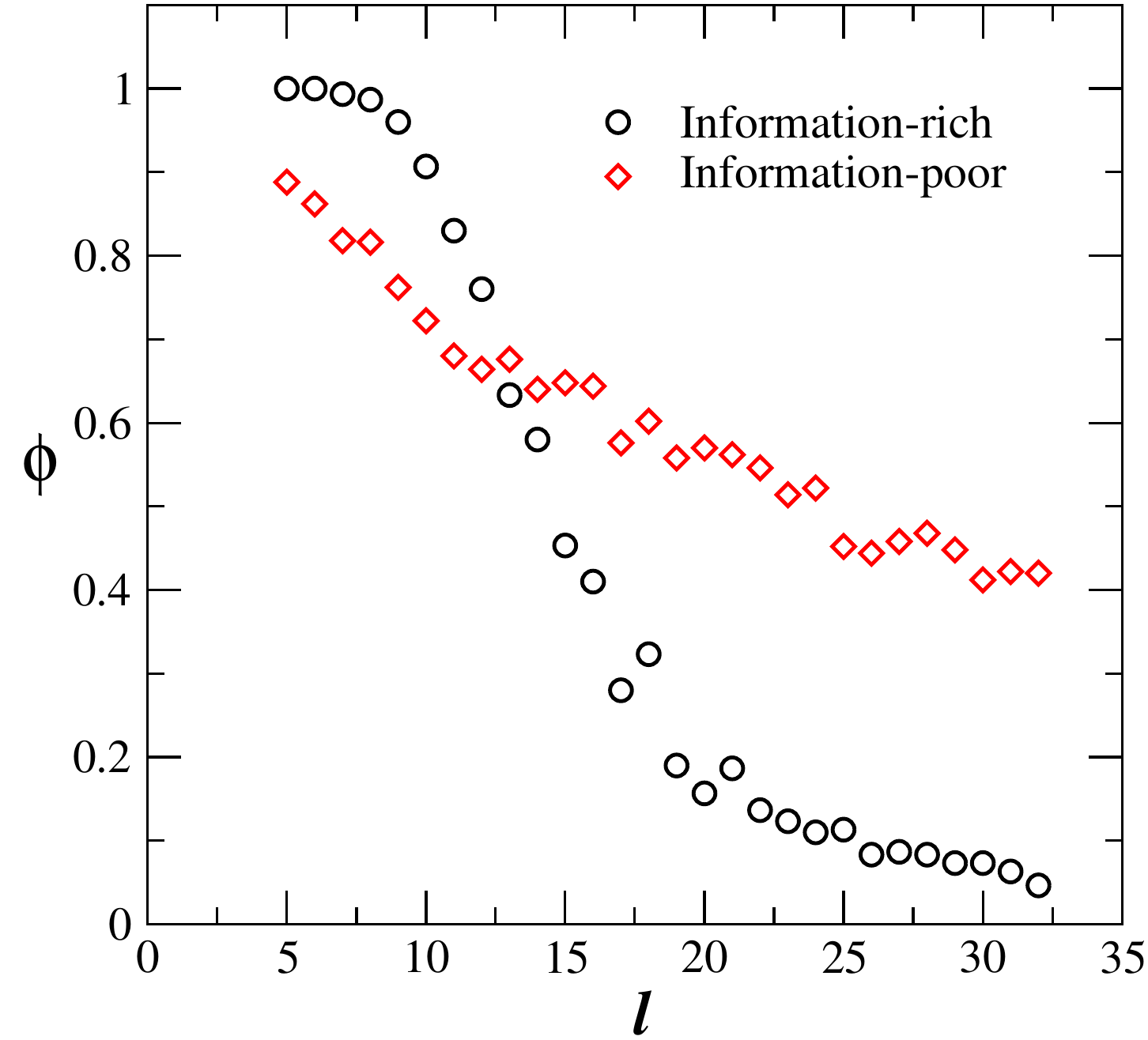}
\caption{\label{funo}Fraction $\phi$ of random (3$^{rd}$ order Markovian) realizations that do
not pass the Lilliefors Gaussianity test for the score distribution versus
motif size $\ell$ (the test is passed by a Gaussian sample).
Averages are over $300$ (information-rich PSFM) and $500$
(information-poor PSFM) samples, respectively.}
\end{center}
\end{figure}

The problem at this point is to establish how significant
$W_{k^\star+1}$ is in statistical terms, {\it i.e.} how unlikely
it is that a particular score has arisen by chance. To this aim,
one normally assumes that scores have a Gaussian distribution and
are uncorrelated along the sequence ({\it i.e.} the $W_k$'s are
independent random variables for different $k$), and evaluates the
likelihood of a given maximum score by employing a Gumbel
distribution\footnote{Recall that the limit cumulative
distribution function of the maximum $M_n$ of a sequence of $n$
independent, identically-distributed random variables is given by
the generalized extreme-value law \beq
\lim_{n\to\infty}\text{Prob}\{M_n\leq x\}:=H_{\xi
}(x)=e^{-\left[1+\xi
x\right]^{-1/\xi}}~~,~~~1+\xi x >0
\label{extreme}\eeq where 
the shape parameter $\xi\in\mathbb{R}$ allows to distinguish three
types of limiting behaviors, depending on whether $\xi>0$
(Fr\`echet), $\xi<0$ (Weibull) or $\xi\to 0$ (Gumbel).}.
Unfortunately, in many cases motifs are short so the number of
terms to be summed up in (\ref{two}) is too small for generating a
Gaussian random variable. The distribution of maxima may thus
deviate significantly from a Gumbel distribution. To appreciate
 how the histogram of scores varies with $\ell$ one can
study the pdf that emerges by applying artificial PSFMs on random
promoters. We have constructed an ensemble of promoters of length
$10000$ using the nucleotide frequencies in the human genome as
the underlying model. On each of these we tested a different
artificial PSFM of size $\ell\times 4$ for
$\ell\in\{5,\ldots,32\}$. We have considered two cases:
information-rich PSFMs, with non-zero entries only for two
(randomly selected) nucleotides for each position;
information-poor PSFMs, which have instead entries drawn from a
uniform distribution on $[0,1/2]$ (the normalization conditions
being obviously enforced). These choices represent limiting cases,
since real data are typically in-between these alternatives.
For each realization we have carried out a Lilliefors test to
probe the normality of the score distribution (other normality
tests such as the Jarque-Bera test return a very similar picture).
Results for the fraction $\phi$ of samples that do not pass the
test are shown in Fig. \ref{funo}.
It is clear that the Gaussian hypothesis is inadequate for short
motifs in both cases. Remarkably, for information-poor PSFMs it is troublesome
also for longer motifs. Note that typical transcription factors BSs
have $6\leq\ell\leq 20$. Clearly, it would be important to outperform
existing computational methods in the presence of information-poor
PSFMs, {\it i.e.} when experimental data on motifs are less sharp.

\section{Accounting for large deviations}

The standard methodology to deal with tail events consists in
selecting a high threshold and studying the exceedances of the
threshold. The basis for this is a theorem by Pickand \cite{pick}.
In simplified terms, it states that given a random variable $X$
and a threshold $u>0$, the distribution function of $Y=X-u$ (the
`excess' over $u$) is such that \beq\label{gdp} \lim_{u\to
x_F}\text{Pr}\{0<Y\leq u\}:=G_{\xi,\sigma}(y)=1-\left(1+\frac{\xi
y}{\sigma}\right)^{-1/\xi} \eeq where $\sigma>0$ is a scale
parameter, $\xi$ is the shape parameter of the distribution of the
maximum value of the random variable $X$ (see footnote 1), and
$x_F$ is the right extremum of the distribution function $F(x)=
\text{Pr}\{X<x\}$, defined by $x_F \equiv \text{inf}\{x:
F(x)=1\}$. $G_{\xi,\sigma}$ is called the generalized Pareto
distribution (GPD). In other words, the GPD is a good
approximation for the distribution of excesses of a random
variable over sufficiently high thresholds. Hence, given the set
of scores $\{W_k\}$ and a threshold $u$, one can obtain estimates
$\ovl{\xi}$ for $\xi$ and $\ovl\sigma$ for $\sigma$ by fitting the
distribution of excesses over $u$ to a GPD. With $\ovl\xi$ and
$\ovl\sigma$ it is possible to evaluate the probability to observe
a score larger than $u$ using (\ref{extreme}). 
Clearly, the smaller is
this quantity, the more significant is the result from a
statistical viewpoint. The parameter estimates will however depend
on the chosen threshold, {\it i.e.} $\ovl\xi\equiv\ovl\xi(u)$ and
$\ovl\sigma\equiv\ovl\sigma(u)$. The problem now consists in
choosing $u$ optimally, so that the condition for the validity of
Pickand's theorem is verified with good accuracy and one still has
enough data above the threshold to be able to estimate the unknown
parameters. As well explained in \cite{embr}, to this aim one can
resort to the following property: let $x_1,\ldots, x_n$ be $n$
independent realizations of a random variable with unknown pdf
$F$, and let \beq e_n(u)=\frac{\sum_{i=1}^n
(x_i-u)\theta(x_i-u)}{\sum_{i=1}^n\theta(x_i-u)} \eeq be the
sample's mean excess over a fixed threshold  $u$, with $\theta(x)$
Heaviside's step function.  Then, if $F$ is a GPD,
 \begin{equation}\label{mean_excess2}
     e_\infty(u) = \frac{\sigma+ \xi u}{1-\xi}
 \end{equation}
This implies that when the empirical plot $e_n(u)$ versus $u$
follows approximately a straight line with a certain derivative
above a value $\ovl u$ of $u$, then the excesses over $\ovl u$
follow approximately a GPD with shape parameter related to the
observed derivative. This allows for an
optimal selection of $\ovl u$ and, in turn, of $\ovl\xi$ and
$\ovl\sigma$.

As said above, once we have estimated these parameters we should
evaluate the statistical significance of the scores via (\ref{gdp}).
This can be accomplished via a Peak-over-Threshold (POT) analysis
\cite{embr,coles}. Consider the excesses of the scores over $\ovl
u$, $Y_k=W_k-\ovl u>0$ (scores that do not exceed $\ovl u$ are
hereafter neglected). Given that the number $N$ of excesses above
a threshold is a Poissonian variable (see {\it e.g.} \cite{hues}),
one easily understands that \bea
\mathcal{P}_k&\equiv&\text{Pr}\left\{\max_{j\in\{0,1,\ldots,L-\ell\}} Y_j>Y_k\right\}\nonumber\\
&=&1-\text{Pr}\left\{\max_{j\in\{0,1,\ldots,L-\ell\}} Y_j\leq Y_k\right\}\nonumber\\
&=&1-\exp\left[-\lambda \left(1 + \ovl\xi
Y_k/\ovl\sigma\right)^{-1/\ovl\xi} \right] \label{otto} \eea
The additional parameter $\lambda$ coming from the Poisson
distribution can be estimated from the data simply as
$\lambda=N/(L-\ell+1)$, where $N$ is the actual number of scores
falling above $\ovl u$ in our sample. Clearly, $\mathcal{P}_k$'s
should be as small as possible for $Y_k$ to be close to the
maximum. A precise condition for real BSs prediction is discussed
below.

\section{Application to the detection of ERE}

We have analyzed a set of 134 promoters whose expression profile
is upmodulated by estrogen, a hormone produced in the ovaries.
Estrogen diffuses across the cell membrane into the cell, where it
interacts with hormones called estrogen receptors. Once activated
by estrogen, receptors act primarily as transcription factors to
regulate the expression of certain genes by binding to DNA.
Estrogen receptors are widely studied in the biomedical literature
since estrogen is related to the development and growth of most
types of breast cancers. Indeed, breast cancer monitoring commonly
includes tests for expression of the estrogen receptor, and
reducing the supply of estrogen is part of breast cancer therapy.
The interaction of an estrogen receptor with DNA occurs at a BS
called estrogen responsive element (ERE, $\ell=13$). The position
of ERE is known experimentally on some promoters but it would be
important to extend this knowledge to other genes that are
sensitive to estrogen. Furthermore, binding at ERE is believed to
be cooperatively linked to binding at two other motifs, called AP2
($\ell=12$) and C/EBP ($\ell=12$). Whether such motifs are present
on all promoters for estrogen-upmodulated genes is however not
known.

We have screened our data set for the (known) presence
of ERE and for the (to be ascertained) presence of AP2 and C/EBP.
The PSFMs for the latter genes have been extracted from the
TransFac database\footnote{Accession numbers: M00189 and M00770.}.
For ERE we have used the PSFM derived in \cite{Cordero}.
 For the sake of clarity, we have subdivided the 134 genes in two groups:
 the first contains the 14 genes for which experimental knowledge is
 available (TFF1, STS, CRKL, NROB2, CYP1B1, FEM1A,
CYP4F11, FOXA1, RPS6KL, NRIP1, CTSD, GAPD, GREB1, IGFBP4)
\cite{ere1, bourdeau}; for the remaining 120 genes information is available
only from computational studies through the NCBI database \cite{NCBI}.

It is now important to discuss the conditions for rejection of a putative motif.
Statistically significant substrings of DNA (indexed $k$) should
satisfy two conditions. On one hand, the value $\mathcal{P}_k^{{\rm (t)}}$ of 
$\mathcal{P}_k$ calculated on the true promoter should be
smaller than a confidence level $\mathcal{P}_c$, since it would be
desirable to minimize the probability of finding a score larger than
$W_k$. $\mathcal{P}_c$ must be chosen so as to guarantee that when
the above procedure is applied to an `engineered' promoter
containing a certain number of motifs, all of these are detected
correctly. In the cases we analyzed, $\mathcal{P}_c$ turns out
to vary in a range between $0.02$ and $0.001$.

Secondly, $\mathcal{P}_k^{{\rm (t)}}$ should be larger than the value
$\mathcal{P}_k^{{\rm (r)}}$ one would obtain when looking for a
real motif on a random promoter, {\it e.g.} one drawn uniformly from
$\{\texttt{A,C,G,T}\}$, with {\it the same} threshold $Y_k$ used for
the real promoter. This condition enforces the expectation that the
score of a certain motif computed on a real DNA sequence should be
higher than that computed on a random string of DNA, where the motif
can only occur by chance. Statistical accuracy can be increased by
considering an ensemble of random promoters rather than just one, and
computing $\mathcal{P}_k^{{\rm (r)}}$ as the average $\mathcal{P}_k$
over the ensemble. Indeed, some random strings will produce larger 
scores than other strings, so it is important to compare the true promoter
directly with the random ones, especially so if the true promoter contains
the motif.

The condition that relevant substrings of length $\ell$ starting
at locus $k+1$ should satisfy is then \beq
\label{nove}\mathcal{P}_k^{{\rm (r)}}<\mathcal{P}_k^{{\rm (t)}}<\mathcal{P}_c
\eeq For comparison, we have considered another condition, less
stringent than (\ref{nove}), namely that both \beq\label{diez}
\mathcal{P}_k^{{\rm(r)}}<\mathcal{P}_c~~~~~
\text{and}~~~~~\mathcal{P}_k^{{\rm (t)}}<\mathcal{P}_c \eeq We shall denote
the latter as the {\it weak} condition and  the former as the {\it strong} one. 
These conditions differ
from the one which is commonly used. Indeed, normally one only looks
for motifs that are unlikely to appear in a random promoter, {\it
i.e.} the only significancy criterion is
$\mathcal{P}_k^{{\rm(r)}}<\mathcal{P}_c$. We show below that our setting ultimately
allows for a reduction in the number of false positives with respect to other
methods, while keeping the same
predictive efficiency ({\it e.g.} the number of true positives) in test cases.

Ultimately, the algorithm we have used to search for ERE, AP2 and
C/EBP on each of the 134 promoters can be summarized as follows.
\begin{enumerate}
\item Define $P_m$ and $P_b$, see (\ref{uno}). The former is given by the experimental PSFMs
of ERE, AP2 and C/EBP. For $P_b$ we have used a 3-step Markovian model
(different choices do not impact results significantly)
\item Compute the scores $\{W_k^{{\rm (t)}}\}_{k=0}^{L-\ell}$ 
for the true promoter and $\{W_k^{{\rm (r)}}\}_{k=0}^{L-\ell}$  
for an ensemble of random promoters generated via a prescribed Markov model 
({\it e.g.} randomly and unformly from $\{\texttt{A,C,G,T}\}$).
\item Estimate the optimal parameters $(\ovl u_{\text{t}},\ovl\xi_{\text{t}},\ovl\sigma_{\text{t}})$ for the true promoter and $(\ovl u_{\text{r}},\ovl\xi_{\text{r}},\ovl\sigma_{\text{r}})$ for the random promoters.
\item Calculate the probability $\mathcal{P}_k^{\text{(t)}}$ for the true promoter, see (\ref{otto}), and $\mathcal{P}_k^{\text{(r)}}$ as the average $\mathcal{P}_k$ over the random promoters.
\item Select motifs with index $k$ satisfying the significancy conditions, either
(\ref{nove}) or (\ref{diez}).
\end{enumerate}
It is worth noting that more than one substring may pass the
significancy tests. In this respect, our choice of computing
$\mathcal{P}_k^{{\rm (t)}}$, that is of considering the likelihood
of observing a particular substring on the real promoter alongside
$\mathcal{P}_k^{{\rm (r)}}$, allows us to draw sharper conclusions
on suboptimal putative motifs since the distribution of the
largest scores on real and random DNA should differ if a motif is
actually present on the real sequence. The fact that more than one
motif may occur obviously doesn't imply cooperation at the
biological level. The method can however be modified to account
for this aspect, {\it e.g.} to identify pairs of correlated motifs
\cite{forth}. 

\section{Results} 
\begin{figure}
\begin{center}
\includegraphics[width=6.5cm]{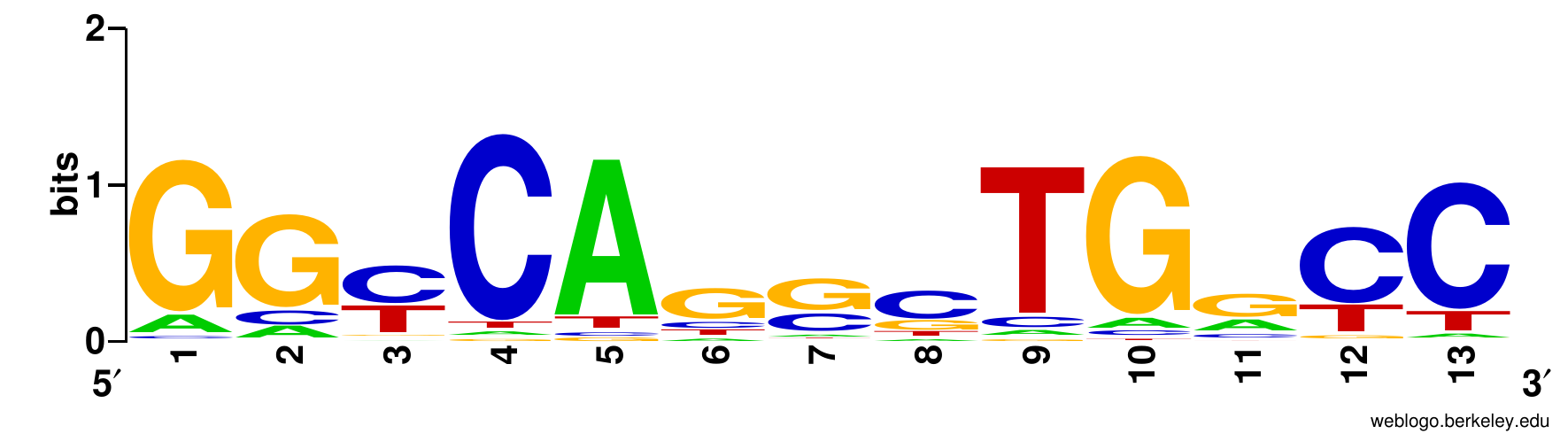}
\includegraphics[width=6.5cm]{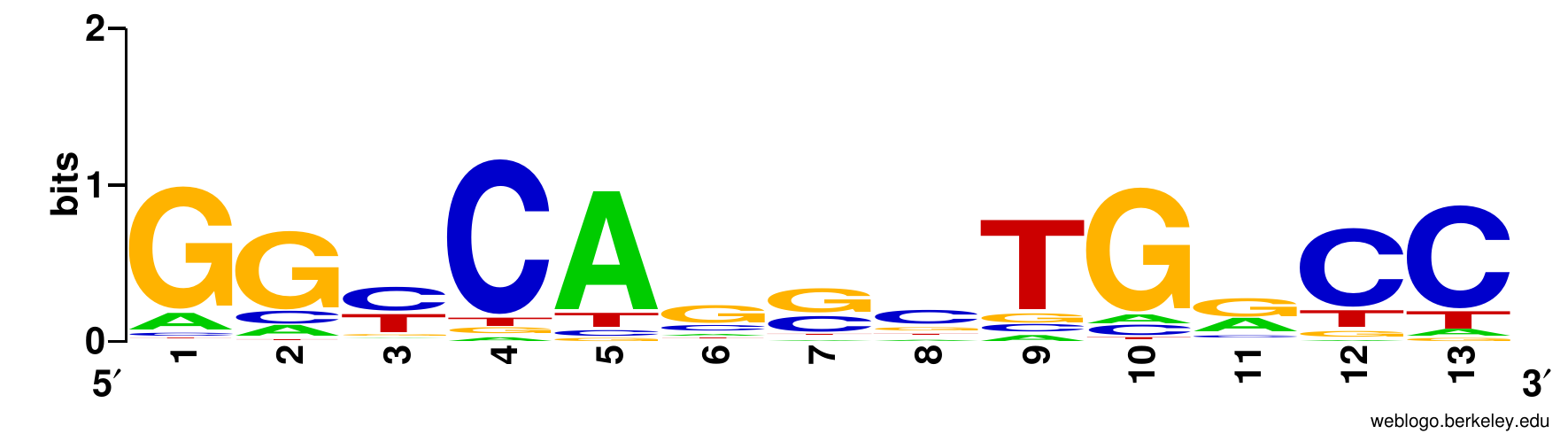}
\caption{\label{fere}Sequence logo of ERE emerging from the 134 promoters
studied. Top: strong significancy condition. Bottom: weak significancy condition.}
\end{center}
\end{figure}

We have detected the presence of ERE on all of the 134
promoters, in agreement with experimental knowledge. In Fig. \ref{fere} we
display the sequence logo\footnote{The frequencies of bases at each position
correspond to the relative heights of letters. The degree of sequence
conservation is instead represented by the total height of a stack of letters,
in units of bits of information.}
 \cite{weblogo} relative to whole data set of
134 promoters.
This should be compared with the sequence
\texttt{GGTCA}$\star\star\star$ \texttt{TGACC} ($\star=$any
nucleotide) constructed by inserting the most frequent nucleotide
in each position and a $\star$ in positions where, experimentally,
every nucleotide can be present. Notice that the sequence is
palindromic, in the sense that the first five bases link to the
last five in reverse order (with the rules \texttt{A-T}, \texttt{C-G}). 
It is clear from Fig. \ref{fere} that our method recovers this property.

\begin{figure}
\begin{center}
\includegraphics[width=7.5cm]{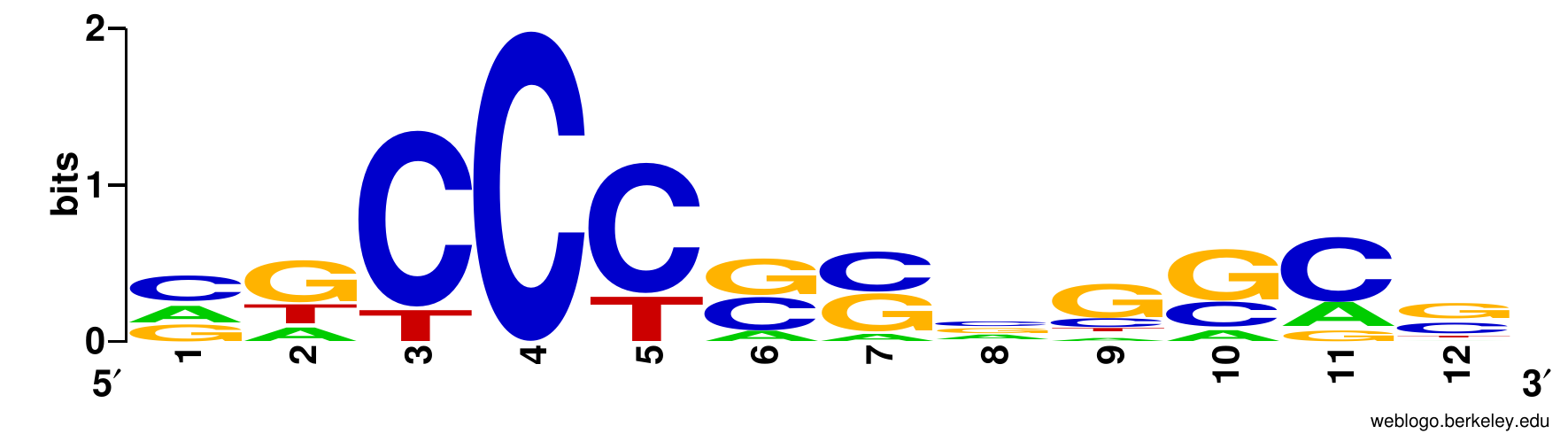}
\includegraphics[width=7.5cm]{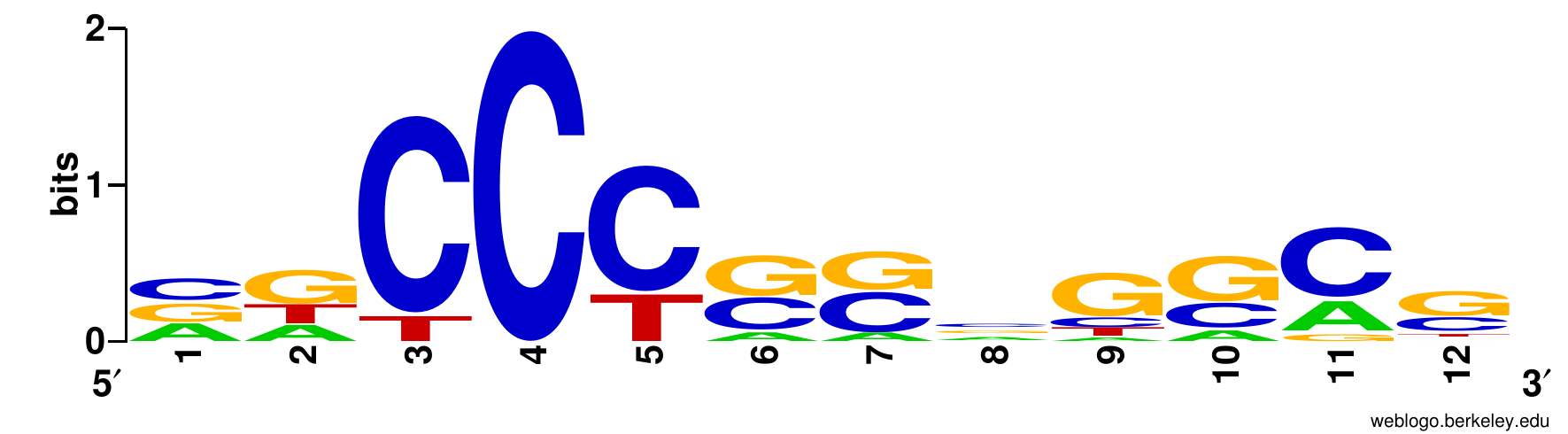}
\caption{\label{fap2}Sequence logo of AP2 emerging from the 134 promoters
studied. Top: strong significancy condition. Bottom: weak significancy condition.}
\end{center}
\end{figure}

\begin{figure}
\begin{center}
\includegraphics[width=7.5cm]{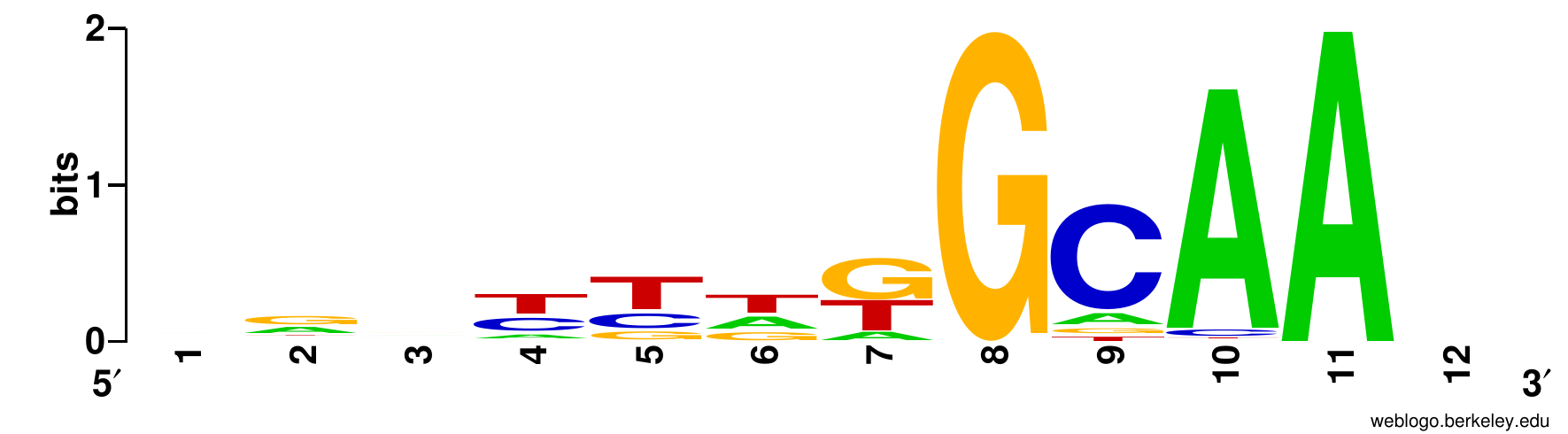}
\includegraphics[width=7.5cm]{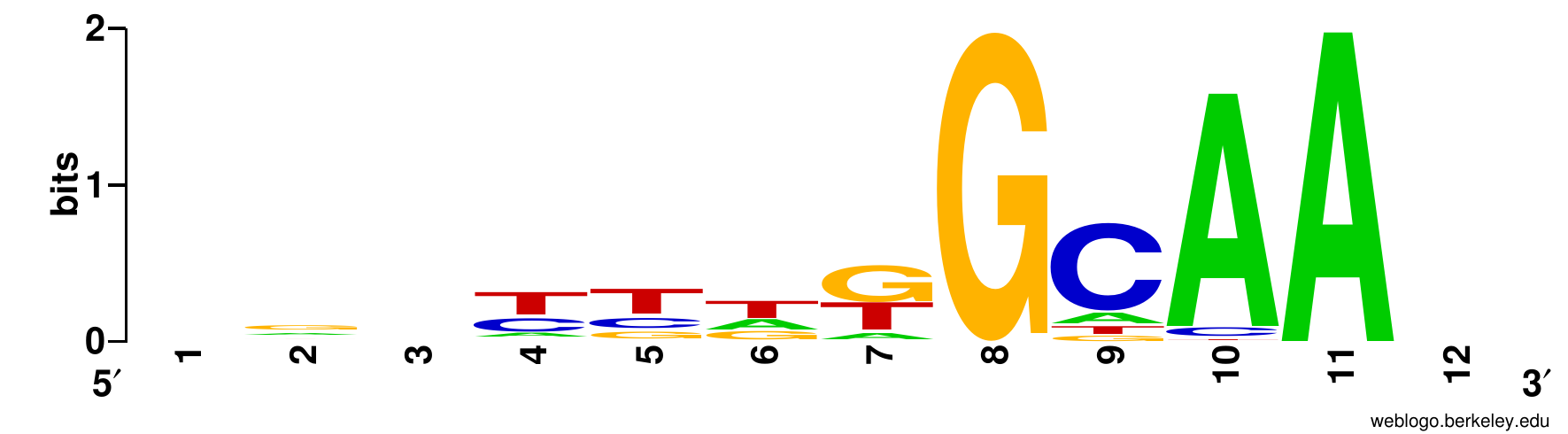}
\caption{\label{cebp}Sequence logo of C/EBP emerging from the 134 promoters
studied. Top: strong significancy condition. Bottom: weak significancy condition.}
\end{center}
\end{figure}

On the contrary the presence of AP2 and C/EBP was not found in
all of the 134 genes (see below for details
from a reduced data set). The resulting sequence logos are shown in
Figures \ref{fap2} and \ref{cebp}.
The former should be compared with the sequence
\texttt{CGCCCGCCGGCG} built with the experimentally
most frequent nucleotides at every position. Note however
that the PSFM for AP2 (from TransFac) includes a small
number of known BSs (13 at the time of writing this article).
For C/EBP, the sequence logo is to be compared to the
experimental highest frequency string \texttt{[G/A]AATTTGGCAAA}, where
the first position is occupied by guanine or adenine with the
same frequency. (In this case a much larger data sample is
available to build the PSFM).

In summary, for the genes we considered the method returns sequences
that are in a very good agreement with the available experimental knowledge
on BSs.

Let us now consider the restricted data set formed by the 14 genes
that have been directly accessed in experiments, at least for ERE.
In Table \ref{tab} we show the outlook of results for the three motifs
we considered.
\begin{table}
\begin{center}
\begin{tabular}{l|c|c|c}
Gene&ERE&AP2&C/EBP\\
\hline
TFF1&Yes&No&Yes\\
STS&Yes&No&No\\
CRKL&Yes&Yes&No\\
NROB2&Yes&No&Yes\\
CYP1B1&Yes&No&No\\
FEM1A&Yes&Yes&Yes\\
CYP4F11&Yes&Yes&No\\
FOXA1&Yes&No&Yes\\
RPS6KL&Yes&No&No\\
NRIP1&Yes&No&Yes\\
CTSD&Yes&Yes&Yes\\
GAPD&Yes&No&No\\
GREB1&Yes&No&No\\
IGFBP4&Yes&No&Yes
\end{tabular}
\caption{\label{tab}Table representing the presence (Yes) or absence (No)
of the ERE, AP2 or C/EBP motif on the genes reported in the first column. Results
are shown for the strong significancy condition.}
\end{center}
\end{table}
With the strong significancy condition, our prediction is that AP2 and
C/EBP are not present on all of the 14 genes, at odds with ERE. An experimental
validation is not yet available.

\begin{figure}
\begin{center}
\includegraphics[width=7.5cm]{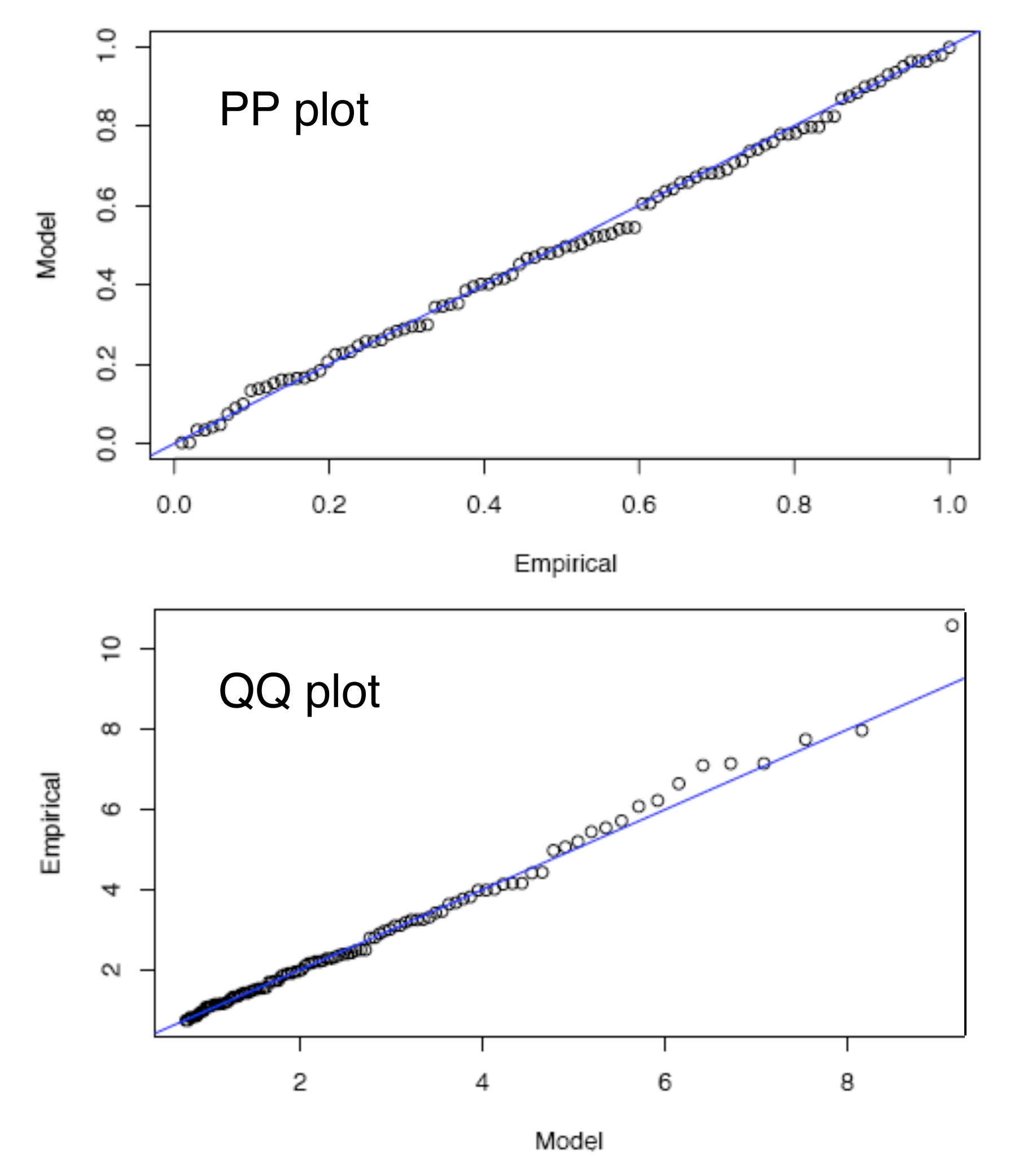}
\caption{\label{QQ}PP-plot (top) and QQ-plot (bottom) for GAPD.}
\end{center}
\end{figure}

Let us now focus on one gene from the data set, namely GAPD (similar
results are obtained for the other genes), and consider ERE. In Fig. \ref{QQ} we display the
probability-probability (PP) and quantile-quantile (QQ) plots for GAPD.
The former shows the empirical probability distribution of excesses versus
a GPD; the latter focuses on the tails, showing the empirical quantiles\footnote{For
a random variable with probability distribution $F(x)$ and for
any $0\leq p\leq 1$, one defines the quantile corresponding to $p$ as
$x(p)=\text{inf}\{x:F(x)\geq p\}$.} of the distribution
of excesses extracted from the data on GAPD versus the quantiles estimated
from a GPD.
These types of plots provide simple measures of plausibility of a certain model.
One sees a convincing agreement between the data and an extreme-value
distribution.

\section{Application to skeletal-muscle specific genes.}

To have an idea of the performance of the method concerning false positives, we have
tested it against a known biological benchmark.
Specifically, we have considered the full set of nine skeletal-muscle specific genes studied in \cite{tomovic}. This set is well studied experimentally. In particular, it is known that six of the transcription
factors from the JASPAR database attach to them \cite{def,tomovic}. The corresponding
motifs have lengths varying from 6 to 12 nucleotides. The best available computational
technique, the Tomovic-Oakeley (TO) method \cite{tomovic}, takes dependencies between sites into account and is able to identify correctly five of the six factors. Table \ref{tab2} compares the performance of our algorithm with that of TO  and with the best available (to our knowledge) algorithm based on cross-species comparison, ConSite \cite{consite}.
\begin{table}[!]
\begin{center}
\begin{tabular}{l|c|c|c}
Gene&ConSite&TO&BT\\
\hline
ALDOA&5/81&5/78&5/70\\
DES&5/80&5/74& 5/70\\
MYOG&5/87&5/85&6/76\\
MYL1& 6/86&5/75&5/71\\
TNNI1&5/81&5/78&5/69\\
MYH7&5/77&4/75&5/76\\
MYH6&5/83&5/78&5/66\\
ACTA1&6/80&5/77&5/67\\
ACTC1&5/84&5/77&5/73
\end{tabular}
\caption{\label{tab2}Comparison between the performance of the ConSite, the Tomovic-Oakeley (TO, including site dependencies) and the present (BT) algorithm on the set of 9 genes studied in \cite{tomovic}. The first number in each entry gives the number of motifs found on the genes (out of 6), the second gives the number of false positives retrieved in the JASPAR database.}
\end{center}
\end{table}
We have chosen our parameters to obtain at least as many true positives as Tomovic-Oakeley. For this setting, the number of false positives is considerably lower in our case.

\section{Conclusions}

Summarizing, we have accounted for large deviations in the distribution of scores for short BSs by a technique that combines well-known properties of extreme-value distributions and a POT analysis. The importance of fluctuations becomes clear if one studies the score distribution in a random setting. This approach allows for a self-consistent estimation of the statistical significance of putative motifs. The general problem of recognizing a small pattern in a large background however presents many open issues. Among these we mention those that have perhaps a more direct biological implication. First, for obvious reasons it would be important to devise methods yielding a still smaller number of false positives. To this aim a deeper analysis of the performance on artificial 
data set would be required, so as to improve the estimation of our parameters and to compare the performances of different methods  on dependence on  $\ell$ and on the structure of the PSFM.  Second, one should address the issue of cooperation between transcription factors. In principle, this requires overcoming the independent-nucleotides approximation and developing techniques that account for score correlations along the sequence. Methods accounting for correlations already exist but they need, at present, more parameters and  larger data set to obtain a reasonable statistical significance. In our case, it is possible to take into account the effect of dependencies by properly grouping scores and applying extreme-value theory to the block scores. This extension is the object of further work \cite{forth}. Clearly, more effective methods would be very welcome and refined statistical and probabilistic tools are likely to play a major role in their development.

\acknowledgments We are deeply indebted with F. Cordero for providing us with the
modified PSFM for ERE, and with R. Calogero for many important discussions and for
a useful collaboration.

\end{document}